\documentclass[12pt]{article}
\usepackage[a4paper]{geometry}
\usepackage{authblk}
\usepackage{amsmath}
\usepackage{supertabular}
\usepackage{subfigure}
\usepackage{graphicx}
\usepackage{epstopdf}
\usepackage[colorlinks=true, citecolor=blue,urlcolor=blue, linkbordercolor={0 0 1}]{hyperref}
\hypersetup{pdfstartview={XYZ null null 1.00}}
\usepackage[none]{hyphenat}
\usepackage{caption}
\usepackage[ruled,vlined]{algorithm2e}
\usepackage{amsmath,amsfonts}
\DeclareMathOperator{\Eq}{Eq}

\title{\bf A note on reducing the computation time for minimum distance and equivalence check of binary linear codes}

\author{\bf Nikolay Yankov}

\affil[1]{\large Faculty of Mathematics and Informatics
	Konstantin Preslavski University of Shumen, 9712 Shumen, Bulgaria, e-mail: \url{jankov_niki@yahoo.com}, ORCID \href{https://orcid.org/0000-0003-3703-5867}{0000-0003-3703-5867}}
\affil[2]{\large College Dobrich, Konstantin Preslavski University of Shumen, 9712 Shumen, Bulgaria, e-mail: \url{kr.enev@shu.bg}}

\author[2]{\bf Krassimir Enev}

\date{}

\begin{document}
\begin{flushleft}
Mathematical and Software Engineering, vol. 4, no. X (2018), 1--3. \\ Var$\varepsilon$psilon Ltd, \ \url{http://varepsilon.com/index.php/mse}
\end{flushleft}
\begingroup
\let\newpage\relax
\maketitle
\endgroup

\begin{abstract} In this paper we show the usability of the Gray code with constant weight words for computing linear combinations of codewords.
This can lead to a big improvement of the computation time for finding the minimum distance of a code.

We have also considered the usefulness of combinatorial $2$-$(t,k,1)$ designs when there are memory limitations to the number of objects (linear codes in particular) that
can be tested for equivalence.
\end{abstract}

\begin{flushleft}
%{\bf Subject Codes:} 68P25, 94A60
\centerline{}
\end{flushleft}

\section{Introduction}
Binary linear codes and self-dual codes in particular are extensively studied for the plethora of connections to communication, cryptography, combinatorial designs, among many.
When computing self-dual codes one should be aware that with the increase of the code length the number of codes also rises exponentially.

The classification of binary self-dual codes begun in 1972 with \cite{Pless1972} wherein all codes of lengths $n \leq 20$ are classified.
Later Pless, Conway and Sloane classify all codes for $n \leq 30$ \cite{Conway1992a}. Next lengths: $32$ is due to Bilous and Van Rees \cite{Bilous2002}, $34$ by Bilous \cite{Bilous2006},
$36$ by Harada and Munemasa in \cite{Harada2010d}. Latest development in this area are for length $38$ in \cite{Bouyuklieva2012} and for $n=40$ due to Bouyukliev, Dzumalieva-Stoeva and Monev in \cite{Bouyukliev2015}.

As length of the code gets bigger the number of codewords rises exponentially and one need efficient algorithms for computing the minimum distance of a linear code, and also efficient ways to check codes for equivalence when there are memory limitations.

This paper is organized as follows: In Section \ref{def} we outline an introduction to linear codes, self-dual codes, combinatorial designs and Gray codes. Next, in Section \ref{s2}, we discuss how a reduction in computation time for minimum distance of linear code with constant-weight Gray code can be achieved. In Section \ref{s3} we explain a method for reducing the computation time for code equivalence by the use of combinatorial 2-designs. We conclude in Section \ref{s4} with a few final notes.

\section{Definitions and preliminaries}\label{def}

Let $\mathbb{F}_q$ be the finite field of $q$ elements, for a prime power $q$.
A linear $[n, k]_q$ \emph{code} $C$ is a $k$-dimensional subspace of $\mathbb{F}_q^n$.
The elements of $C$ are called \emph{codewords},
and the \emph{(Hamming) weight} of a codeword $v\in C$ is the number of the non-zero coordinates of $v$.
We use $\text{wt}(v)$ to denote the weight of a codeword.
The \emph{minimum weight} $d$ of $C$ is the minimum nonzero weight of any codeword in $C$
and the code is called an $[n, k, d]_q$ code. A matrix whose rows form a basis of $C$ is called
a \emph{generator matrix} of this code.% (denoted by $\gen(C)$).

Let $(u,v)\in\mathbb{F}_q$ for $u,v\in\mathbb{F}_q^n$ be an inner
product in $\mathbb{F}_q^n$. The \emph{dual code} of an $[n,k]_q$
code $C$ is $C^{\perp}=\{u \in \mathbb{F}_q^n \mid (u,v)=0$ for
all $v \in C \}$ and $C^{\perp}$ is a linear $[n,n-k]_q$ code.
In the binary case the inner product is the standard one, namely, $(u,v)=\sum_{i=1}^n{u_iv_i}.$
If $C
\subseteq C^{\perp}$, $C$ is termed \emph{self-orthogonal}, and if
$C = C^{\perp}$, $C$ is \emph{self-dual}. We say that two binary linear codes $C$ and $C'$ are
\emph{equivalent} if there is a permutation of coordinates which sends $C$ to $C'$.
In the above definition the code equivalence is an equivalence relation is a binary relation that is reflexive, symmetric and transitive.
Denote by $\Eq(a,b)$ some function that checks for equivalence all pairs of elements in both sets of linear codes $a$ and $b$.
For more information on codes we encourage the reader to \cite{Huffman2003}.

When working with linear codes it is often needed for certain algorithm to pass trough all (or part) of binary vectors of given length.
One way to make the generation efficient is to ensure that successive elements are generated such that they differ in a small, pre-specified way. One of the earliest examples of such a process is the Gray code generation. Introduced in a pulse code communication system  in 1953 \cite{Gray1953}, Gray codes now have applications in diverse areas: analogue-to-digital conversion, coding theory, switching networks, and more. For the past 70 years Gray codes have been extensively studied and currently there are many different types of Gray code.

\emph{A binary Gray code} of order $n$ is a list of all $2^n$ vectors of length $n$ such that exactly one bit changes from one string to the
next.

A $t$-$(v, k, \lambda)$ design $D$ is a set $X$ of $v$ points together with a collection of $k$-subsets of
$X$ (named blocks) such that every $t$-subset of $X$ is contained exactly in $\lambda$ blocks. The block
intersection numbers of $D$ are the cardinalities of the intersections of any two distinct blocks.

\section{Reducing computation time for minimum distance of linear code with constant-weight Gray code}\label{s2}

Assume we have a linear binary $[n,k]$ code $\cal C$ and we need to find its minimum distance $d.$ Denote
by $\cal G$ the generator matrix of the code $\cal C$ with rows $r_1,\ldots,r_k.$ The obvious and direct approach is to compute all codewords of $\cal C$ and find their weight. This means that all $2^k$ linear combinations of $t$ $(1\leq t\leq k)$ of the rows of $\cal G$ must be computed using Algorithm \ref{Alg1}.

\begin{algorithm}[H]
  \caption{The direct approach}\label{Alg1}
\begin{verbatim}
for (i1 = 1; i1 <= k-t+1; i1++) {
    for (i2 = i1+1; i2 <= k-t+2; i2++) {
      for (i3 = i2+1; i3 <= k-t+3; i3++) {
        ...
          for (it = itm1+1; it <= k; it++) {body}... }}
\end{verbatim}
$\ $\\[-8mm]
\end{algorithm}

Then for each of the $\binom{k}{t}$ combination we need to compute $t$ cycles and essentially $t$ operations. Furthermore, in the body of this
algorithm we need to find the codeword $c\in{\cal C}$ which is a linear combination of those rows of the generator matrix $\cal G$ that are chosen for the current combination, i.e. $c=\sum\limits_{s=1}^{t}r_{i_s},$ which will be represented by $t$ ``exclusive or'' (\texttt{xor}) operations $c=r_{i_1}\oplus r_{i_2}\oplus \ldots \oplus r_{i_t}$.

Our approach is to use Gray code for generating combinations in such a way that each successive combination is generated
by the previous one with only two \texttt{xor} operations. Two \texttt{xor} operations are the absolute minimum
since, if we have to switch from one combination of $t$ elements to another, one \texttt{xor} will add or remove a position making a $t+1$ or a $t-1$ combination.
In \cite{Tang1973} it was proved that the set of $\binom{k}{t}$-vectors of weight $t,$ when
chained according to the ordering on the Gray code ${\cal G}_k,$ has a Hamming distance of exactly two between every
pair of adjacent code vectors. Also in \cite{Tang1973} an algorithm for generating the constant-weight code vectors on a Gray code was given.
Later in \cite{Bitner1976} a more efficient recursive algorithm was introduced (Algorithm \ref{Alg2}).

\begin{algorithm}[!htb]
  \caption{Constant $t$-weight $(0<t\leq k)$ Gray code ${\cal G}_t$ \cite{Bitner1976}}\label{Alg2}
\textbf{for} $j=1$ \textbf{to} $t$ \textbf{do} $\left\{\begin{array}{ll}
g_j=1\\
\tau_j=j+1
\end{array}\right.$

\textbf{for} $j=t+1$ \textbf{to} $k+1$ \textbf{do} $\left\{\begin{array}{ll}
g_j=0\\
\tau_j=j+1
\end{array}\right.$

$s=k$

$\tau_1=k+1$

$i=0$

\textbf{while} $i{<}k{+}1$ \textbf{do }
$\left\{\begin{array}{@{}l}
\text{\textbf{output}\ }(g_k,g_{k-1}\ldots, g_1)\\
i=\tau_1\\
\tau_1=\tau_i\\
\tau_i=i+1\\
\text{\textbf{if}\ }g_i=1
\text{\ \textbf{then}}\left\{
\begin{array}{@{}l}
\text{\textbf{if\ }}s\neq0\text{\textbf{\ then\ }}g_s=\overline{g}_s\\
\ \ \ \ \ \ \ \ \ \ \ \ \text{\textbf{else\ }}g_{i-1}=\overline{g}_{i-1}\\
s=s+1
\end{array}
\right.\\
\ \ \ \ \ \ \ \ \ \ \ \ \ \text{\textbf{else}}\left\{
\begin{array}{@{}l}
\text{\textbf{if\ }}s \neq1 \text{\textbf{\ then\ }}g_{s-1}=\overline{g}_{s-1}\\
\ \ \ \ \ \ \ \ \ \ \ \ \text{\textbf{else\ }}g_{i-1}=\overline{g}_{i-1}\\
s=s-1
\end{array}
\right.\\
g_i=\overline{g}_i\\
\text{\textbf{if}\ } s=i-1\text{\ or\ }s=0\\
\ \ \ \ \ \ \ \ \text{\textbf{then}\ }s=s+1\\
\ \ \ \ \ \ \ \ \text{\textbf{else}}\left\{
\begin{array}{@{}l}
s=s-g_{i-1}\\
\tau_{i-1}=\tau_i\\
\text{\textbf{if\ }}s=0\text{\textbf{\ then\ }}\tau_1=i-1\\
\ \ \ \ \ \ \ \ \ \ \ \ \text{\textbf{else\ }}\tau_1=i+1\\
\end{array}
\right.\\
\end{array}
\right.$

\end{algorithm}

What we want to do is to find in Gray code ${\cal G}_k$ those $k$-tuples that have the same weight $t$, for example when $k=4$ for $t=1$ we have:
0000$\rightarrow$\textbf{0001}$\rightarrow$0011$\rightarrow$\textbf{0010}$\rightarrow$0110$\rightarrow$0111 $\rightarrow$0101$\rightarrow$\textbf{0100}$\rightarrow$1100$\rightarrow$1101$\rightarrow$1111$\rightarrow$1110$\rightarrow$1010$\rightarrow$1011$\rightarrow$1001$\rightarrow$\textbf{1000}
and similarly, for $t=2$ we have:
0000$\rightarrow$0001$\rightarrow$0011$\rightarrow$0010$\rightarrow$\textbf{0110}$\rightarrow$0111$\rightarrow$\textbf{0101}$\rightarrow$0100$\rightarrow$\textbf{1100}$\rightarrow$1101$\rightarrow$ 1111$\rightarrow$1110$\rightarrow$\textbf{1010} $\rightarrow$1011$\rightarrow$ \textbf{1001}$\rightarrow$1000. Note that Algorithm \ref{Alg2} starts with the word $1^t0^{k-t}$ and finishes with $0^{k-t}1^t.$

\textbf{Example 1:} If we need to find all triples in ${\cal G}_6$ we have a total of 20 triples. We start with 000111 and from Gray code we have the following sequence of positions to change
\begin{align*}
&[2,4],[1,2],[1,3],[2,5],[1,2],[2,3],[1,4],[1,2],[1,3],[2,6],[1,2],[2,3],\\
&[3,4],[1,5],[1,2],[2,3],[1,4],[1,2],[1,3].
\end{align*}
%111000$\rightarrow$101100$\rightarrow$011100$\rightarrow$110100$\rightarrow$100110$\rightarrow$
%010110$\rightarrow$001110$\rightarrow$101010$\rightarrow$011010$\rightarrow$110010$\rightarrow$
%100011$\rightarrow$010011$\rightarrow$001011$\rightarrow$000111$\rightarrow$100101$\rightarrow$
%010101$\rightarrow$001101$\rightarrow$101001$\rightarrow$011001$\rightarrow$110001

So the sequence of triples is as follows
\begin{align*}
&\{1,2,3\}, \{1,3,4\}, \{2,3,4\}, \{1,2,4\},\{1,4,5\},\{2,4,5\}, \{3,4,5\}, \{1,3,5\}, \{2,3,5\}, \{1,2,5\}, \\ &\{1,5,6\},\{2,5,6\},\{3,5,6\},\{4,5,6\},\{1,4,6\},\{2,4,6\},\{3,4,6\},\{1,3,6\},\{2,3,6\},\{1,2,6\}.
\end{align*}

Usually, when we need to compute the minimum weight of a binary code $C,$ we start with the initializing $r_1\oplus\cdots\oplus r_t,$ then
we need the pair of position that should be changed to obtain the next $t$-tiple and so on. Since for given $i (1<i<2^t)$ it is easy to find the $i$-th $t$-weight vector and begin with the linear combination generated by it, the algorithm can be parallelized to accommodate its use on multiple CPU cores.

\section{Reducing computation time for code equivalence with combinatorial 2-designs}\label{s3}

What can be done when there are more linear codes that the equivalence algorithm can accommodate in the allowed memory.
We consider the case when all codes have the same weight enumerator and also the same order of their automorphism group. This means that all other options for reducing the number of codes we are considering are exhausted.

The question then is: How can we efficiently ensure that the algorithm will check every pair of codes.
If we have $s\in\mathbb{N}$ times more codes that that algorithm can check, we can split this into $2s$ halves of sets of codes and then
check all $\binom{2s}{2}=s(2s-1)$ pairs for equivalence. This is not very efficient since this has the quadratic $O(s^2)$ efficiency.
The more efficient way is to use $2$-$(v,k,1)$ combinatorial design, which ensures that every pair of points (sets of codes in our case) appear exactly in one block and is checked for equivalence only once. Such designs exists, for example, when $\lambda=1$ and $v=k^2,$ we have a projective plane: $X$ is the point set of the plane and the blocks are the lines \cite{Tonchev2017}.

\begin{figure}[!htb]
  % Requires \usepackage{graphicx}
\begin{center}
  \includegraphics[width=5cm]{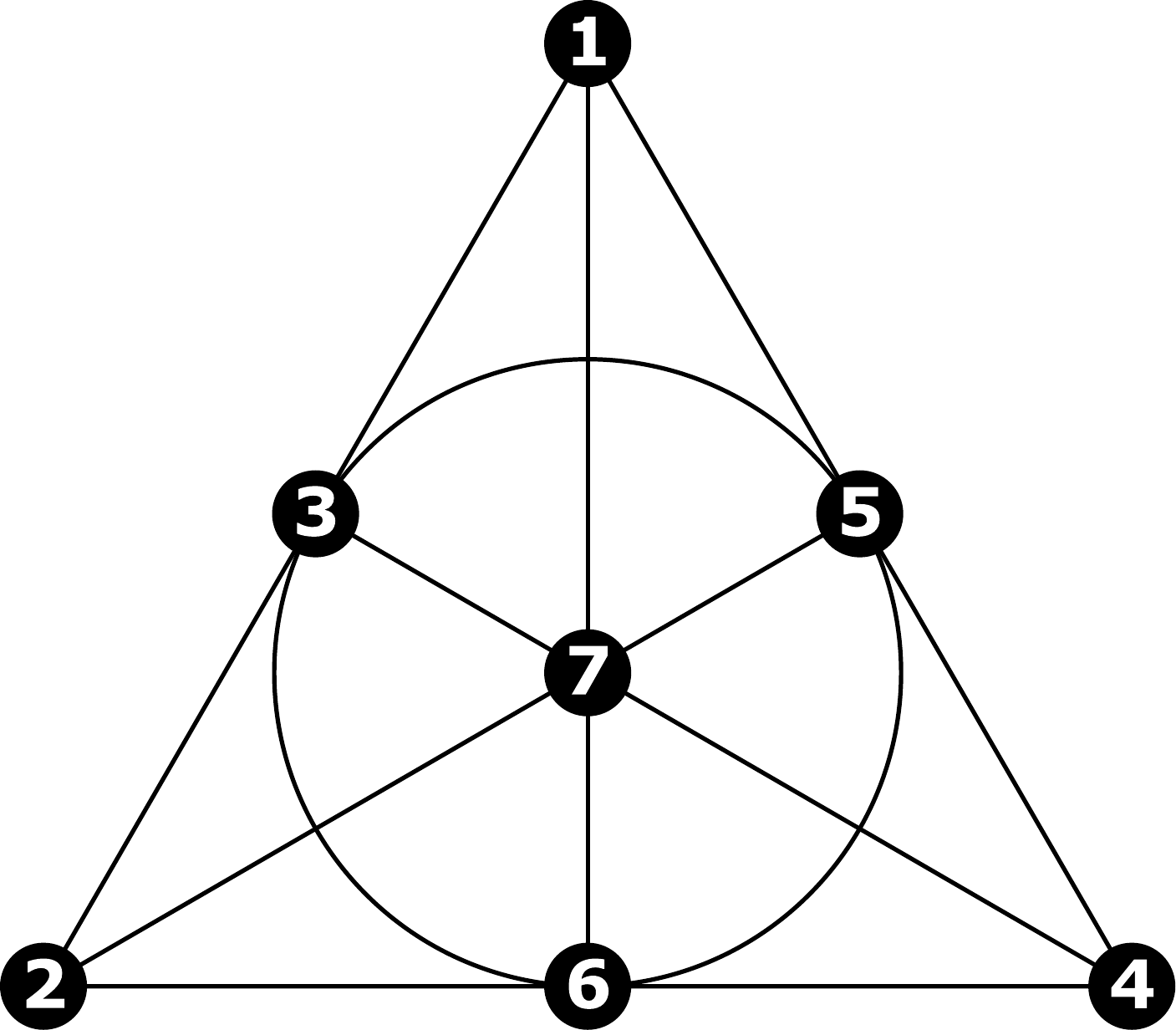}
\end{center}
  \caption{Fano plane}\label{fano}
\end{figure}

For example, consider the case of 7 sets $i_1,\dots,i_7$ of binary self-dual codes.
If we use the standard approach we should do the tests $\Eq(i_j,i_s),$ $1\leq i_k<i_j\leq 7$ for all $\binom{7}{2}=21$ pairs of sets. Now, consider using the combinatorial design approach, viz. the Fano plane (see \cite{CRC}) illustrated in Fig. \ref{fano}. It is well known that the Fano plane is a combinatorial $2$-$(7,3,1)$-design \cite{CRC}. This means that every pair of sets $(i_j,i_s),$ $1\leq i_j<i_s\leq 7$ appear in exactly one of the 7 blocks (the blocks of Fano plane are the 6 lines and the circle), so if a code is present in different sets it is reduced to only one copy.

Using the ordering of the sets $i_j\prec i_s$ iff $j<s,$ we can use the following sequence for automorphism testing:
$$\Eq(i_1{,}i_2{,}i_3){,}\Eq(i_1'{,}i_4{,}i_5){,}\Eq(i_1',i_6,i_7){,}\Eq(i_2',i_4',i_6'){,}\Eq(i_2',i_5',i_7'){,}\Eq(i_3',i_4'',i_7''){,}\Eq(i_3',i_5'',i_6''),$$
where $i_j'$ means that the interval $i_j$ is purged of the codes that are equivalent to codes from preceding sets,
$i_j''$ means that the interval $i_j'$ is purged of the codes that are equivalent to codes from preceding sets, and so on.
As a result the reduced inequivalent set of codes will be the union $i_1'\cup i_2'\cup i_3'\cup i_4'''\cup i_5'''\cup i_6'''\cup i_7'''.$

\section{Conclusions}\label{s4}

In the present research we have considered the usability of the Gray code with constant weight words for computing linear combinations of codewords.
We have shown that, in this way, a big improvement of the computation time for finding the minimum distance of a code can be achieved.

We have also considered the usefulness of combinatorial $2$-$(t,k,1)$ designs when there are memory limitations to the number of objects (linear codes in particular) that can be tested for equivalence. In our example we have shown that using the Fano plane one can achieve complete classification with as much as half of the computation time needed otherwise. It remains to find efficient designs for different number of sets to be checked for equivalence.

\subsubsection*{Acknowledgement}
The authors express their gratitude to prof. Borislav Panayotov for the invitation to publish in this journal.
This work was supported by European Regional Development Fund and the Operational Program ``Science and Education for Smart Growth'' under contract UNITe No BG05M2OP 001-1.001-0004-C01 (2018-2023).

{\footnotesize Copyright $\copyright$ 2018 First Author. This is an open access article distributed under the Creative Commons Attribution License, which permits unrestricted use, distribution, and reproduction in any medium, provided the original work is properly cited.}

\end{document}